\documentclass[pra, showpacs, twocolumn, a4paper, nofootinbib]{revtex4}
\usepackage{graphicx}
\usepackage{amsmath, amsfonts, amssymb, bm}
\usepackage{amsmath}
\usepackage{verbatim}
\usepackage{bm}

\begin{document}

\newcommand{\ket}[1]{| #1 \rangle} 
\newcommand{\bra}[1]{\langle #1 |}

\title{Delayed sudden birth of entanglement} 
\author{Zbigniew \surname{Ficek}$^{a}$}
\email{ficek@physics.uq.edu.au}
\author{Ryszard \surname{Tana\'{s}}$^{b}$}
\affiliation{$^{a}$Department of Physics, The University of Queensland, Brisbane, Australia 4072\\ 
  $^{b}$Nonlinear Optics Division, Institute of Physics, Adam
  Mickiewicz University, Pozna\'n, Poland}

\date{\today}

\begin{abstract}
  The concept of time delayed creation of entanglement by the dissipative process of spontaneous emission 
  is investigated. A threshold effect for the creation of entanglement is found that the initially unentangled 
  qubits can be entangled after a finite time despite the fact that the coherence between the qubits exists for 
  all times. This delayed creation of entanglement, that we 
  call sudden birth of entanglement, is opposite to the currently extensively discussed sudden death of 
  entanglement and is characteristic for transient dynamics of one-photon entangled states of the system. 
  We determine the threshold time for the creation of entanglement and find that it is 
  related to time at which the antisymmetric state remains the only excited state being populated.
  It is shown that the threshold time can be controlled by the distance between the qubits and the direction 
  of initial excitation relative to the interatomic axis. This effect suggests a new alternative for the study 
  of entanglement and provides an interesting resource for creation on demand of entanglement between two qubits.
\end{abstract}


\maketitle


Dynamical creation of entanglement in the presence of a noisy environment and its disentangled properties 
are problems of fundamental interest in quantum computation and quantum information processing. They have 
attracted a great deal of attention especially in connection with the phenomenon of decoherence induced by 
spontaneous emission resulting from the interaction with the environment that leads to irreversible loss of 
information encoded in the internal states of the system and thus is regarded as the main obstacle in practical 
implementations of entanglement. Contrary to intuition that spontaneous emission should have a destructive effect 
on entanglement, it has been shown that under certain circumstances this irreversible process can in fact entangle 
initially unentangled qubits~\cite{beige}, thus implying a kind of quantum coherence induced in the emission. 
This effect has been studied for identical qubits coupled to a common multi-mode vacuum field or coupled to a damped 
single-mode cavity field and has a simple explanation in terms of the collective nature of the spontaneous emission 
from a system of qubits being located within a transition wavelength of each other or coupled to a single-mode cavity field.
In this terminology, the system can be represented in terms of the collective (Dicke) symmetric and antisymmetric 
states that decay with significantly different rates~\cite{dic,le70}. Both states are maximally entangled states, but 
the entanglement results solely from the trapping properties of the antisymmetric state of the system. More precisely, 
with the initially only one qubit excited, 
a part of the initial population is trapped in the antisymmetric state from which it cannot decay or may decay 
much slower than the populations of the remaining states. That is the reason, why the system decays to an entangled 
long living mixed state involving only the antisymmetric and the ground states of the system. In this way, an 
entanglement persisted over a long time is obtained dynamically via spontaneous emission. The degree of the  
entanglement such generated is determined by the population of the antisymmetric state that with the initially 
only one atom excited approaches a steady state value of one-half.

Apart from the constructive effect of spontaneous emission on entanglement, it has been shown that some 
entangled states of two qubits can have interesting decoherence properties that two initially entangled 
qubits can reach separability abruptly in a finite time that is much shorter than the exponential decoherence 
time of spontaneous emission~\cite{sd}. This drastic non-asymptotic feature of entanglement 
has been termed as the "entanglement sudden death", and is characteristic of the dynamics of a special class of 
initial two-photon entangled states. 
In fact, the effect shows up only if specific initial two-photon coherences are created between the qubits.
A recent experiment of Almeida {\it et al.}~\cite{al07} with correlated $H$ and $V$-polarized photons has shown 
evidence of the sudden death of entanglement under the influence of independent environments. The required initial 
two-photon coherence was created by the parametric down-conversion process.
Considering the present interest in understanding of the decoherence in entangled qubits, it presents 
a fascinating example of a dynamical process in which spontaneous emission affects entanglement and coherences
in very different ways. Although the sudden death feature is concerned with the disentangled properties of spontaneous 
emission there can be interesting "sudden" features in the temporal creation of entanglement from initially 
independent qubits. If such features exist, they would provide an interesting resource for creation on demand
of entanglement between two qubits.
 
In this paper we show that a "sudden" feature in the temporal creation of entanglement exists in a dissipative time 
evolution of interacting qubits. We term this feature as delayed (sudden) birth of entanglement, as it is opposite to 
the sudden death of entanglement, and show that the feature arises dynamically with initially separable qubits.
The delayed creation of entanglement is not found in the small sample Dicke model that ignores the 
evolution of the antisymmetric state. It is also not found in a system with initially only one qubit excited.
 For this, an initial entanglement and the experimentally difficult individual addressing 
of qubits are not required. The initial conditions considered here include both qubits inverted that can be done using
a standard technique of a short $\pi$ pulse excitation. The second initial condition considered here involves excitation 
by a short $\pi/2$ pulse which leaves qubits separable but simultaneously prepared in the superposition of their energy states.
We carry our considerations in the context of concurrence and two-level atoms interacting through the vacuum field 
and analyse how the concurrence evolves in time. We determine the threshold time for creation of 
entanglement and discuss the dependence of the magnitude of the entanglement on the distance between the qubits and 
direction of excitation relative to the inter-qubit axis.
Related calculations have appeared involving entanglement creation via spontaneous emission~\cite{beige}.
However, these calculations studied a limited set of initial conditions and as such these calculations miss the novel
feature of delayed birth of entanglement that depends on specific initial conditions of both qubits. The sudden birth
of entanglement deserves more careful study, especially in view of its fundamental importance in a controled creation 
of entanglement on demand in the presence of a dissipative environment.

The usual way to identify entanglement between two qubits in a mixed state is to examine the concurrence, an entanglement 
measure that relates enlangled properties to the coherence properties of the qubits~\cite{woo}. For a system described by 
the density matrix $\rho$, the concurrence $C$ is defined as 
\begin{eqnarray}
  {\cal C}(t) =
  \max\left(0,\lambda_{1}(t)-\lambda_{2}(t)-\lambda_{3}(t)
    -\lambda_{4}(t)\,\right) ,\label{e1} 
\end{eqnarray}
where $\{\lambda_{i}(t)\}$ are the square roots of the eigenvalues of the non-Hermitian matrix $\rho(t)\tilde{\rho}(t)$ 
with
\begin{equation}
  \tilde{\rho}(t) = \sigma_{y}\otimes\sigma_{y}\,\rho^{*}(t)\,\sigma_{y}\otimes\sigma_{y} ,\label{e2}  
\end{equation}
and $\sigma_{y}$ is the Pauli matrix. The range of concurrence is from 0 to 1. For unentangled (separated) 
atoms ${\cal C}(t)=0$, whereas ${\cal C}(t)=1$ for the maximally entangled atoms.

The density matrix, which is needed to compute ${\cal C}(t)$ and written in the basis of the separable product states
$\ket 1 =\ket{g_{1}g_{2}}, \ket 2 =\ket{e_{1}g_{2}}, \ket 3 =\ket{g_{1}e_{2}}, \ket 4 =\ket{e_{1}e_{2}}$ is in general 
composed of sixteen nonzero density matrix elements. 
However, in the case of the simple dissipative evolution of the system without any initial coherences between 
the qubits and without the presence of coherent excitations, the density matrix takes a simple block diagonal form
\begin{eqnarray}
  \rho(t) = \left(
    \begin{array}{cccc}
      \rho_{11}(t) & 0 & 0 & 0 \\
      0 & \rho_{22}(t) & \rho_{23}(t) & 0\\
      0 & \rho_{32}(t) & \rho_{33}(t) & 0\\
      0 & 0 & 0 &\rho_{44}(t)
    \end{array}\right) ,\label{e3}
\end{eqnarray}
in which we put all the coherences, except the one-photon coherences $\rho_{23}(t)$ and $\rho_{32}(t)$, equal to zero.
As we will see the zeroth coherences remain zero for all time, that they cannot be created by spontaneous decay. However, 
the coherences $\rho_{23}(t)$ and $\rho_{32}(t)$ can be created by spontaneous emission even if they are initially zero.

For a system described by the density matrix (\ref{e3}), the concurrence has a simple analytical form
\begin{eqnarray}
  {\cal C}(t) = \max\left\{0,\,\tilde{{\cal C}}(t)\right\} ,\label{e4} 
\end{eqnarray}
with
\begin{eqnarray}
  \tilde{{\cal C}}(t) = 2|\rho_{23}(t)| -2\,\sqrt{\rho_{11}(t)\rho_{44}(t)} . \label{e5} 
\end{eqnarray}
It is evident that there is a threshold for the coherence at which the system becomes entangled. Thus, the non-zero 
coherence $\rho_{23}(t)$ is the necessary condition for entanglement, but not in general sufficient one since there 
is also a rather subtle condition of a minimum coherence between the qubits. 

Alternatively, we may study conditions for entanglement by writing the concurrence (\ref{e5}) in terms of the 
maximally entangled Dicke symmetric $\ket s = (\ket 2 +\ket 3 )/\sqrt{2}$ and antisymmetric $\ket a = (\ket 2 -\ket 3 )/\sqrt{2}$ 
states
\begin{eqnarray}
  \tilde{{\cal C}}(t) &=& \sqrt{\left(\rho_{ss}(t)-\rho_{aa}(t)\right)^{2} - \left(\rho_{sa}(t)-\rho_{as}(t)\right)^{2}} \nonumber \\
   && -2\,\sqrt{\rho_{11}(t)\rho_{44}(t)} , \label{e6} 
\end{eqnarray}
which shows that the threshold for entanglement depends on the distribution of the population between the 
entangled and separable states. Notice that the threshold depends on the population of the upper state $\ket 4$. 
Thus, no threshold features can be observed in entanglement creation by spontaneous emission for qubits initially prepared 
in a single photon state.

In addition to the threshold phenomenon, there is also an evident competition between the symmetric and antisymmetric 
states in creation of entanglement. We see that the best for creation of entanglement through the one-photon states 
is to populate either symmetric or antisymmetric states but not both simultaneously. Thus, one could expect that a large 
entanglement can be created when one of the two entangled states is excluded from the dynamics and remains 
unpopulated for all times.

However, we demonstrate a somehow surprising result that entanglement cannot be created by spontaneous emission 
in the Dicke model that excludes the dynamics of the antisymmetric state.  The Dicke model 
 is composed of three states in a ladder configuration: the upper state $\ket{4}$, the intermediate symmetric state $\ket{s}$
and the ground state $\ket{1}$. Physically, the Dicke model represents two qubits confined to a region much smaller than the 
resonant wavelength~\cite{dic}. In this case, the time evolution of the density matrix elements under the spontaneous emission
is determined by the following density matrix elements~\cite{le70}
\begin{eqnarray}
  \rho_{44}(t) &=& \rho_{44}(0)\,{\rm e}^{-2\gamma t}  ,\nonumber \\
  \rho_{ss}(t) &=& \rho_{ss}(0)\,{\rm e}^{ -2\gamma t} + 2\gamma t \rho_{44}(0)\,{\rm e}^{-2\gamma t} ,\nonumber \\
  \rho_{aa}(t) &=& \rho_{aa}(0)   ,\label{e7a}
\end{eqnarray}
which shows that the antisymmetric state does not participate in the spontaneous dynamics of the system. The 
population of the antisymmetric state remains constant in time. As a result, if the system is prepared in the antisymmetric state
it stays there for all times. We are, however, interested in the dynamical creation of entanglement by spontaneous emission
from separable states to entangled states.

\begin{figure}[th]
  \includegraphics[height=4.2cm]{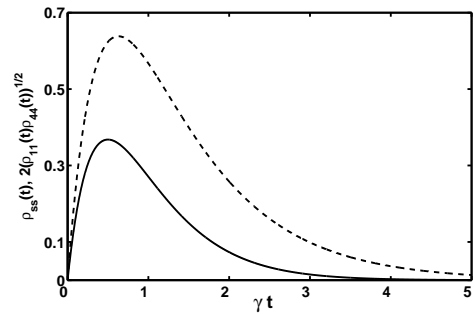}
  \caption{The time evolution of the population $\rho_{ss}(t)$ (solid line) and the threshold factor 
  $2\sqrt{\rho_{11}(t)\rho_{44}(t)}$ (dashed line) 
  for initially both qubits inverted,  $\rho_{44}(0)=1$.}
  \label{fig1a}
\end{figure}

In the Dicke model, the only entangled state that participates in the spontaneous dynamics is the symmetric state $\ket{s}$, so 
let us see if one can create entanglement by spontaneous emission that can populate the symmetric state 
from the upper state $\ket{4}$. 
Figure~\ref{fig1a} shows the time evolution of the population $\rho_{ss}(t)$ and the threshold factor $2\sqrt{\rho_{11}(t)\rho_{44}(t)}$ 
for the initially fully inverted qubits. We see that the threshold factor overweights the population $\rho_{ss}(t)$ for all times, which indicates that despite of a large population of the symmetric state, no entanglement is created. Thus, we may conclude that 
spontaneous emission cannot create entanglement in the Dicke model where only the symmetric state participates in the atomic dynamics. 

We now turn to the system of two qubits that are separated by distances comparable to the resonant wavelength. In this
case,  the antisymmetric states fully participates in the spontaneous dynamics and the time evolution of the density matrix 
elements under the spontaneous emission and for an arbitrary initial state is given by~\cite{le70}
\begin{eqnarray}
  \rho_{44}(t) &=& \rho_{44}(0)\,{\rm e}^{-2\gamma t}  ,\nonumber \\
  \rho_{ss}(t) &=& \rho_{ss}(0)\,{\rm e}^{-(\gamma+\gamma_{12}) t} \nonumber \\
   &+& \rho_{44}(0)\,{\rm e}^{-2\gamma t}\frac{\gamma +\gamma_{12}}{\gamma -\gamma_{12}}
  \left({\rm e}^{(\gamma-\gamma_{12}) t} -1\right) ,\nonumber \\
  \rho_{aa}(t) &=& \rho_{aa}(0)\,{\rm e}^{-(\gamma-\gamma_{12}) t} \nonumber \\ 
   &+& \rho_{44}(0)\,{\rm e}^{-2\gamma t}\frac{\gamma -\gamma_{12}}{\gamma +\gamma_{12}}
  \left({\rm e}^{(\gamma +\gamma_{12})t} -1\right)   ,\nonumber \\
  \rho_{sa}(t) &=& \rho_{sa}(0)\,{\rm e}^{-(\gamma +2i\Omega_{12})t}   ,\label{e7}
\end{eqnarray}
and $\rho_{11}(t)= 1- \rho_{44}(t)-\rho_{ss}(t)-\rho_{aa}(t)$.
Note that the full solution for the density matrix elements exhibits the effect of the cooperative damping $\gamma_{12}$ 
and the dipole-dipole interaction $\Omega_{12}$.

We consider spontaneous creation of entanglement in the system initially prepared in a separable state. 
The entanglement depends, of course, on the initial state of the system. We consider two examples of initial states.
As the first example,  consider a state which covers a broad class of initial states in which the qubits are prepared in 
the superposition of their energy states
\begin{eqnarray}
   \ket{\Psi_{0}} = \frac{1}{2}\left(\ket{g_{1}}
   +i{\rm e}^{i\vec{k}\cdot \vec{r}_{1}}\ket{e_{1}} \right)\otimes 
\left(\ket{g_{2}} +i{\rm e}^{i\vec{k}\cdot \vec{r}_{2}}\ket{e_{2}} \right), \label{e8} 
\end{eqnarray}
where $\vec{k}$ is the wave vector of the excitation field.
The initial state $\ket{\Psi_{0}}$ is separable and can be created in practice by an incident~$\pi/2$ pulse excitation of each qubit.
In this case, the initial values of the density matrix elements~are
\begin{eqnarray}
   \rho_{sa}(0) &=& \frac{1}{4}i\sin\vec{k}\cdot\vec{r}_{12} ,\quad 
   \rho_{ss}(0)  = \frac{1}{4}\left( 1 +\cos\vec{k}\cdot\vec{r}_{12}\right) ,\nonumber \\
   \rho_{aa}(0) &=& \frac{1}{4}\left( 1 -\cos\vec{k}\cdot\vec{r}_{12}\right) ,\quad
   \rho_{44}(0)  = \frac{1}{4}, \label{e9} 
\end{eqnarray}
which shows that a particular initial state depends on the distance between the qubits and the direction of excitation relative to 
the interatomic axis.
\begin{figure}[th]
  \includegraphics[height=5.0cm]{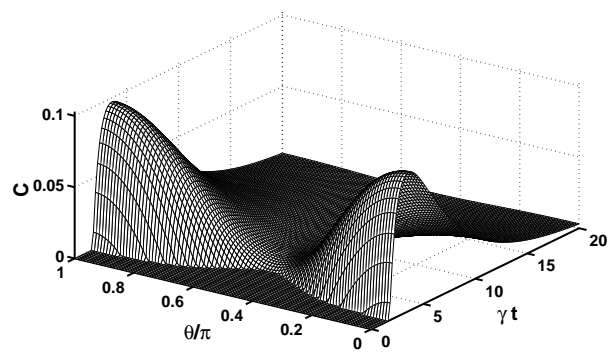}
  \caption{The time evolution of the concurrence and its dependence on the direction of excitation relative to 
  the inter-atomic axis for $r_{12}/\lambda =0.25$ and the polarization of the atomic dipole moments
$\vec{\mu}\parallel \vec{r}_{12}$.}
  \label{fig2}
\end{figure}

Figure~\ref{fig2} shows the concurrence as a function of time and the angle $\theta$ between the excitation direction 
and the vector $r_{12}$ connecting the atoms. It is seen that there is no entanglement at earlier times independent of  the direction
of excitation, and {\it suddenly} at some finite time an entanglement starts to build up. However, no entanglement builds up 
if the system is initially excited in the direction perpendicular to the interatomic axis. One can see from Eq.~(\ref{e9}) that 
in this case the system is excited through the symmetric state. Thus, similar to the Dicke model, entanglement in the system
cannot be created by an excitation of the system through the symmetric state. A large entanglement is created only if the system 
is excited in the direction of the interatomic axis. This means that crucial for entanglement creation by spontaneous emission 
is an excitation through the antisymmetric state.

\begin{figure}[th]
  \includegraphics[height=4.2cm]{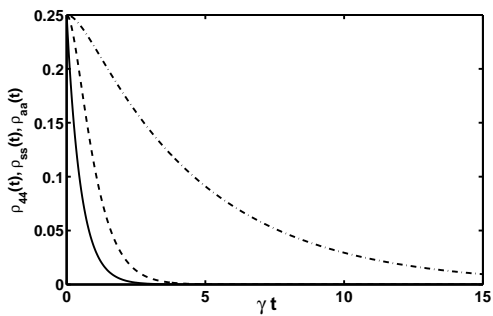}
  \caption{The time evolution of the populations $\rho_{44}(t)$ (solid line), $\rho_{ss}(t)$ (dashed line) and 
  $\rho_{aa}(t)$ 
  (dash-dotted line) for $\theta =0$, $r_{12}/\lambda =0.25$ and $\vec{\mu}\parallel \vec{r}_{12}$.}
  \label{fig3}
\end{figure}

The above conclusion is supported by the analysis of the time evolution of the population of the excited states of the system
that is illustrated in Fig.~\ref{fig3}. It is quite evident from the figure that at the time $t\approx 4/\gamma$ when the entanglement 
starts to build up, the antisymmetric state is the only excited state of the system being populated. This effect is attributed to 
the slow decay rate of the antisymmetric state. The state decays on the time scale of $(\gamma -\gamma_{12})^{-1}$ that is 
much shorter than the decay time of the symmetric and the upper states.

In the second example, we consider the qubits initially prepared in their excited states, that can be realized in practice 
by a short $\pi$ pulse excitation. In this case 
\begin{eqnarray}
   \rho_{44}(0)  = 1 ,\quad \rho_{sa}(0)=\rho_{ss}(0)=\rho_{aa}(0)=0 . \label{e10} 
\end{eqnarray}

\begin{figure}[th]
  \includegraphics[height=6cm]{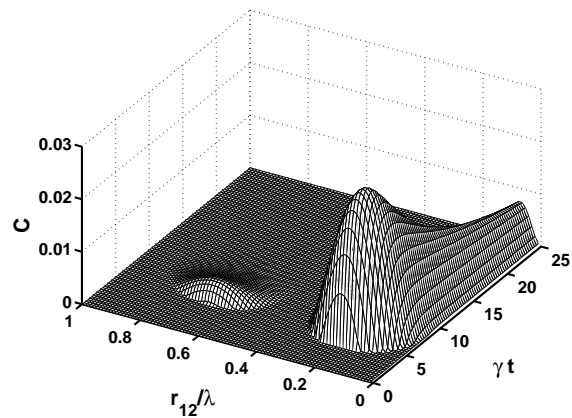}
  \caption{The time evolution of the concurrence and its dependence on the distance between two, initially inverted
  qubits.}
  \label{fig1}
\end{figure}

Figure~\ref{fig1} illustrate the concurrence as a function of time and the distance between the qubits. Similar to the first 
example, illustrated in Fig.~\ref{fig2},  there is no entanglemnet at earlier times, but suddenly at some finite time an entanglement starts to build up. However, it happens only for a limited range of the distances $r_{12}$. It is easy to show that 
the "islands" of entanglement seen in Fig.~\ref{fig1} appear at distances for which $\gamma_{12}$ is different 
from zero.
One can easily show that similar to the first example, the entanglement seen in Fig.~\ref{fig1} decays out on a time 
scale $(\gamma -\gamma_{12})^{-1}$ that is the time scale of the population decay from the antisymmetric state.


In summary, we have predicted an interesting phenomenon of delayed (sudden) birth of entanglement 
that initially separable qubits become entangled via spontaneous emission after a finite time.
In contrast to the sudden death phenomenon that involves two-photon entangled states, the sudden
birth involves one-photon entangled symmetric and antisymmetric states. 
We have demonstrated that the participation of the antisymmetric state in the dynamics is crucial 
for creation of entanglement in the systems.


This work was supported by the Australian Research Council and the University 
of Queensland Travel Awards for International Collaborative Research.

\end{document}